\documentclass[letterpaper,twocolumn,10pt]{article}
\usepackage{usenix}

\usepackage{tikz}
\usepackage{amsmath}

\usepackage{xcolor}
\usepackage{tcolorbox}
\usepackage[table]{xcolor}
\newtcolorbox{titleEnv}{
colframe=black!80,
colback=gray!10,
fonttitle=\bfseries,
coltitle=black,
left=3pt,
right=3pt,
top=3pt,
bottom=3pt,
boxrule=0.4mm,
arc=3mm
}

\usepackage{siunitx}

\usepackage{enumitem}
\usepackage{graphicx}
\usepackage{booktabs} 
\usepackage{multicol}
\usepackage{multirow}
\usepackage{appendix}
\usepackage{titlesec}
\usepackage{setspace}
\usepackage{makecell}
\usepackage{url}
\usepackage{booktabs}
\usepackage{multirow}
\usepackage{array}
\usepackage{makecell}
\usepackage{graphicx}
\usepackage{xcolor}

\definecolor{ruleblue}{RGB}{32,92,148}
\definecolor{rulebg}{RGB}{247,250,253}
\definecolor{ruleframe}{RGB}{190,210,230}

\newcommand{\lcf}{\textcolor{black}}

\usepackage{array} 
\usepackage{amsmath}
\usepackage{booktabs}
\usepackage{makecell}
\usepackage{array}
\usepackage{xcolor}

\usepackage{booktabs}
\usepackage{makecell}
\usepackage{array}
\usepackage{pifont}
\usepackage{booktabs}
\usepackage{tabularx}
\usepackage{array}
\usepackage{makecell}
\usepackage{amsmath}
\usepackage{stfloats}

\newcolumntype{L}[1]{>{\raggedright\arraybackslash}p{#1}}
\newcolumntype{Y}{>{\raggedright\arraybackslash}X}
\usepackage{xcolor}

\definecolor{ruleblue}{RGB}{32,92,148}
\definecolor{rulebg}{RGB}{247,250,253}
\definecolor{ruleframe}{RGB}{190,210,230}

\definecolor{ruleblue}{RGB}{31, 92, 150}
\definecolor{rulebg}{RGB}{247, 250, 253}
\definecolor{ruleframe}{RGB}{184, 207, 231}

\newtcolorbox{rulecard}[2][]{
  enhanced,
  breakable,
  colback=rulebg,
  colframe=ruleframe,
  boxrule=0.45pt,
  leftrule=2.2pt,
  arc=1.5pt,
  left=6pt,
  right=6pt,
  top=5pt,
  bottom=5pt,
  fonttitle=\bfseries,
  coltitle=black,
  title={#2},
  attach boxed title to top left={xshift=4pt,yshift=-2pt},
  boxed title style={
    colback=white,
    colframe=ruleframe,
    boxrule=0.4pt,
    arc=1.5pt
  },
  #1
}

 \tcbuselibrary{skins}

\begin{document}

\date{}

\title{From Component Snapshots to Lifecycle Traces: Agent-Based Software Composition Analysis}

\author{
Chaofan Li,
Zhengduo Xue,
Chengxiang Li,
Yutao Hu,
Yueming Wu,
Deqing Zou
\\
Huazhong University of Science and Technology, China
}

\maketitle

\begin{abstract}
Software supply-chain security requires accurate identification of third-party components and an understanding of how they evolve from development to execution. Existing \emph{software composition analysis} (SCA) approaches examine manifests, build environments, release artifacts, containers, or runtime states, but typically produce only stage-specific views of software composition. As dependencies are resolved, removed, repackaged, and transformed across lifecycle stages, a single snapshot cannot capture both where a component originates and where it ultimately ends up. Combining snapshots from multiple stages still leaves their cross-stage relationships unresolved. We present \emph{SCA-Agent}, an agent-based approach to lifecycle-aware SCA that reconstructs evidence-backed component lifecycle traces across Code, Build, Release, Deploy, and Runtime. \emph{SCA-Agent} adaptively explores project-specific analysis paths, gathers stage-specific evidence, and correlates observations across stages to recover component identities, versions, introduction paths, propagation relationships, and final lifecycle states. We evaluate \emph{SCA-Agent} on 105 real-world projects from the Java, JavaScript, and Python ecosystems. \emph{SCA-Agent} achieves the highest component detection F1 across all lifecycle stages and ecosystems. For vulnerability exposure assessment, it reaches an F1 score of 96.69\%, exceeding the best traditional SCA tool by 18.76 percentage points. These results show that lifecycle-aware SCA supports traceable component provenance and more accurate software supply-chain risk assessment.
\end{abstract}

\section{Introduction}
\label{sec:introduction}

\textbf{Background.}
Modern software systems increasingly rely on third-party components and open-source dependencies, making the software supply chain an integral part of software development and delivery \cite{ohm2020backstabber,torres2019toto,openssf_slsa}. Components are incorporated into downstream software through direct or transitive dependencies, allowing vulnerabilities to propagate along dependency relationships and ultimately affect applications. To identify third-party components and their dependencies, \emph{software composition analysis} (SCA) and \emph{software bills of materials} (SBOMs) have been widely adopted for component identification, vulnerability detection, and supply-chain auditing \cite{ponta2018beyond,10.1145/3597503.3623347,souppaya2022secure}.

\noindent \textbf{Existing Approach: Snapshot-oriented SCA.}
Existing SCA techniques typically analyze a particular software representation or execution state, such as manifests and lockfiles in source code, resolved dependencies \cite{10646983} in the build environment \cite{ponta2020detection,lin2025context,yu2024correctness}, release artifacts \cite{alia2026memsbom}, container images \cite{Kawaguchi2024UnderstandingTE}, or runtime loading information \cite{ponta2020detection,lin2025context}. Although these approaches use different detection techniques, they follow the same analysis paradigm: given a software state, they recover the composition observable in that state. Their outputs are therefore \emph{component snapshots}, capturing the components and dependency relationships visible at a particular stage.

\noindent \textbf{Limitation of Snapshot-oriented SCA.}
Software composition does not remain static throughout software development and delivery \cite{9794026}. As software moves from source code through build, release, deployment, and execution, components may be resolved, introduced, pruned, repackaged, or transformed, changing both the resulting composition and the security-relevant information available for analysis. We therefore focus on five lifecycle stages directly associated with such composition evolution: Code, Build, Release, Deploy, and Runtime. These transformations can produce substantially different component sets across stages \cite{10.1145/3689944.3696164}. For example, Druid v1.2.28 contains 102, 522, 76, 78, and 58 components at the Code, Build, Release, Deploy, and Runtime stages, respectively \cite{alibaba_druid_1228}. Consequently, a snapshot at any single stage cannot capture how software composition evolves throughout the lifecycle.

Different snapshots also preserve complementary security information. Earlier stages typically retain dependency origins and introduction paths, whereas later stages provide stronger evidence about which downstream states a component reaches~\cite{cofano2024sbom,shu2025tool}. At Build, dependency resolution can recover direct and transitive dependency chains and identify which upstream dependency introduces a component~\cite{zhao2023software,na2024cneps,schwaighofer2023extending}. Yet Build alone cannot determine whether that component survives packaging and reaches release or deployment~\cite{cofano2024sbom,shu2025tool}. Runtime observations provide stronger evidence that a component participates in an execution~\cite{cofano2026transparent}, but they are workload-dependent and often lack the dependency context required to trace the component back to its introducer~\cite{rasheed2026hidden,cofano2026transparent}. Hence, no single snapshot can simultaneously determine a component's \emph{Origin}, where it is introduced, and its \emph{Fate}, which downstream lifecycle stages it reaches.

\noindent \textbf{Beyond Multiple Snapshots.}
A natural alternative is to analyze multiple lifecycle stages \cite{foo2019dynamics}. Simply collecting multiple snapshots still leaves their observations disconnected. Components may undergo version resolution, repackaging, and representation transformations, so source dependencies, build dependencies, artifact files, and runtime modules do not have a natural one-to-one correspondence \cite{rasheed2026hidden}. Connecting a component's Origin to its Fate therefore requires recovering how it is carried, transformed, or eliminated as software transitions across stages. We refer to these cross-stage relationships as component \emph{Propagation}. Propagation enables two security-relevant analyses: determining whether an early-stage vulnerable dependency reaches downstream software, and tracing a vulnerable component observed at a later stage back to the dependency that introduced it. Thus, the fundamental limitation of snapshot-oriented SCA is not merely incomplete coverage, but the absence of cross-stage relationships that connect component provenance with its downstream state.

\noindent \textbf{Key Insight.}
We therefore extend the unit of analysis in SCA from independent \emph{component snapshots} to \emph{component lifecycle traces}. A lifecycle trace connects a component's \emph{Origin} to its \emph{Fate} through evidence-backed \emph{Propagation} relationships. It captures where a component is introduced, how it propagates across lifecycle stages, and which downstream states it reaches. This representation supports both forward reasoning about whether a dependency reaches downstream software and backward tracing from a later-stage component to its introducer.

\noindent \textbf{Main Challenges.}
Reconstructing component lifecycle traces presents two key challenges. First, lifecycle analysis paths are project-specific and progressively discovered during execution. Projects differ substantially in build systems, module structures, artifact formats, and deployment mechanisms, while later analysis targets often depend on objects discovered in earlier steps. Second, lifecycle evidence must be associated across heterogeneous software representations. Manifest dependencies, resolved dependencies, artifacts, and runtime modules expose different forms of evidence \cite{dietrich2024security}, while component versions and representations may change across stages. Accurate lifecycle reconstruction therefore requires both adaptive lifecycle exploration and cross-stage reasoning over distributed evidence.

\noindent \textbf{Our Approach.}
To reconstruct lifecycle traces under project-specific and dynamically discovered analysis paths, we design and implement \emph{SCA-Agent}, an agent-based system for lifecycle-aware software composition analysis. \emph{SCA-Agent} adaptively explores lifecycle stages and analysis objects according to the target project's structure and intermediate observations. It collects stage-specific evidence across Code, Build, Release, Deploy, and Runtime, then performs cross-stage reasoning to resolve component identities, versions, and propagation relationships. By connecting these observations, \emph{SCA-Agent} reconstructs evidence-backed lifecycle traces describing each component's \emph{Origin}, \emph{Propagation}, and \emph{Fate}, and incorporates them into a lifecycle-aware SBOM.

\noindent \textbf{Evaluation.}
We evaluate \emph{SCA-Agent} on 105 real-world open-source projects across Java, JavaScript, and Python. Our analysis confirms substantial component evolution across lifecycle stages. \emph{SCA-Agent} reconstructs component lifecycle traces with Origin Accuracy ranging from 84.44\% to 99.11\% and Fate Accuracy from 70.94\% to 97.95\% across the three ecosystems. For lifecycle-aware vulnerability exposure assessment, it achieves an F1 score of 96.69\%, outperforming the best traditional SCA tool by 18.76 percentage points. Ablation studies further show that adaptive planning and cross-stage reasoning are important for lifecycle reconstruction.

\noindent \textbf{Contributions.}
The contributions are summarized as follows:

\noindent $\bullet$ We propose \emph{Lifecycle-aware SCA}, extending software composition analysis from isolated component snapshots to lifecycle traces that connect component \emph{Origin}, \emph{Propagation}, and \emph{Fate}.

\noindent $\bullet$ We design and implement \emph{SCA-Agent}, which combines adaptive lifecycle exploration, multi-stage evidence collection, and cross-stage reasoning to reconstruct evidence-backed component lifecycle traces.

\noindent $\bullet$ We systematically evaluate \emph{SCA-Agent} on 105 real-world open-source projects and demonstrate its effectiveness in lifecycle reconstruction and lifecycle-aware vulnerability exposure assessment.
\section{Motivation and Conceptual Model}

\subsection{Motivating Example}
\label{fig:R2}
\begin{figure}[t]
    \centering
    \includegraphics[width=1\linewidth]{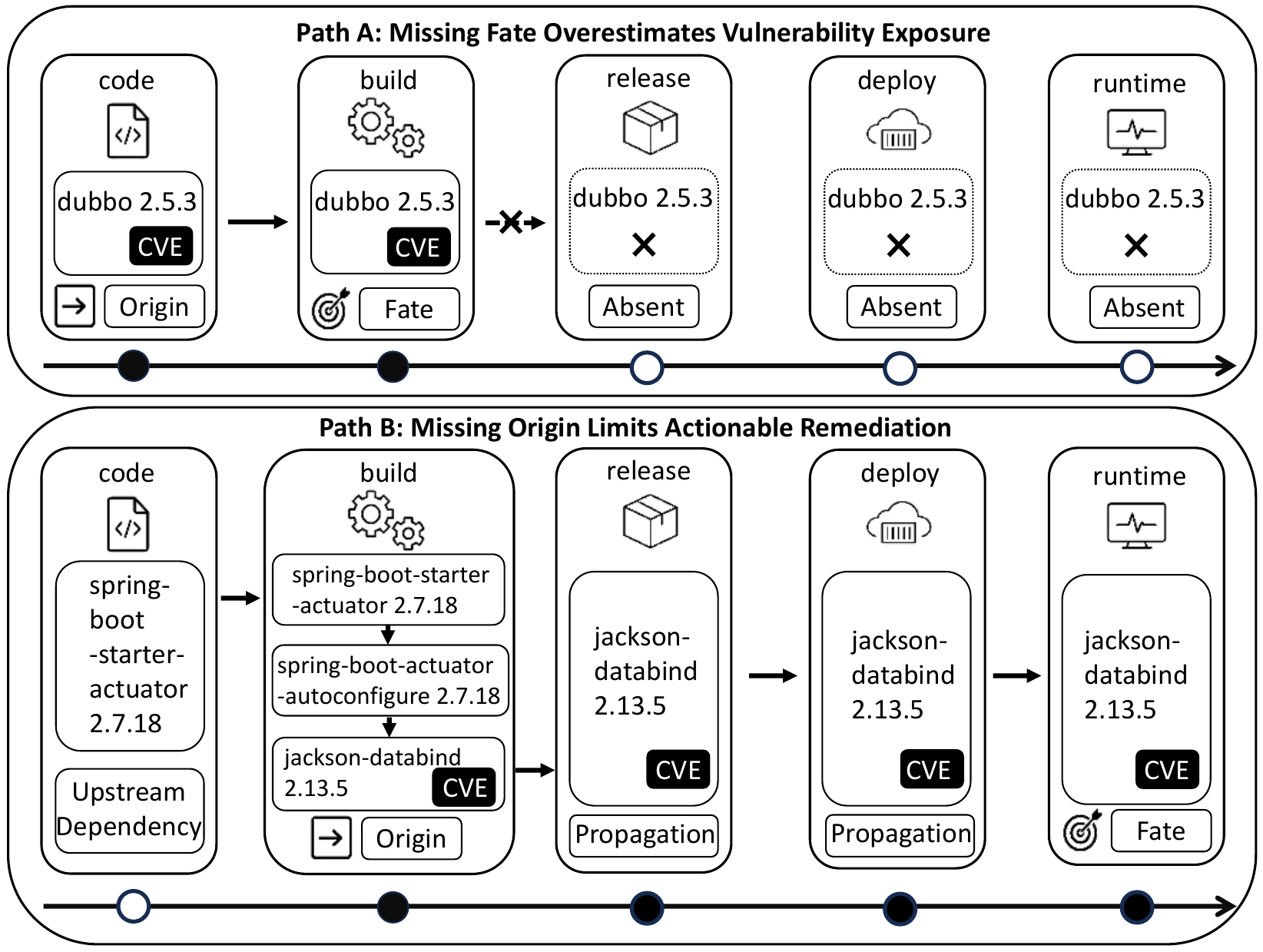}
    \caption{Motivating Example}
    \label{fig:R2}
\end{figure}

To illustrate the limitations of snapshot-based software composition analysis in software supply-chain security assessment, we take a real‑world project incorporating \textsc{Druid}-v1.2.28 as our motivating example to examine how vulnerable components manifest across diverse software states, as shown in Figure~\ref{fig:R2}. Existing SCA tools typically generate component snapshots from a particular class of software objects, such as resolved dependencies, release artifacts, or runtime environments, and perform vulnerability detection based on these snapshots. However, identifying a component in a single snapshot does not fully characterize its security impact on the final software~\cite{bifolco2025dependency}.

\noindent \textbf{Path A: Missing Fate Overestimates Vulnerability Exposure.}
In the component snapshot generated at the Build stage, dependency resolution identifies {com.alibaba:dubbo@2.5.3} together with its associated vulnerabilities, causing snapshot-based vulnerability detection to report it as a potential risk. However, examining subsequent software states reveals that dubbo@2.5.3 does not appear in the Release artifact, Deploy environment, or Runtime state. In other words, the Build snapshot can confirm the presence of the vulnerable component in the build dependency graph, but cannot determine whether it ultimately reaches the delivered software. This is not an isolated case: among the {173 CVE} matched at the Build stage of this project, {94 (54.3\%)} correspond to vulnerable components that do not appear in the Release artifact. This demonstrates that {detecting a vulnerability in a build snapshot does not necessarily imply that the vulnerability is ultimately exposed in the delivered software}.

\noindent  \textbf{Path B: Missing Origin Limits Actionable Remediation.}
A different limitation arises for vulnerable components that do reach the final software. In this project, \lcf{{com.fasterxml.jackson.core:jackson-databind@2.13.5} }is a transitive dependency that persists across subsequent software states, allowing Release, Deploy, or Runtime snapshots to confirm that the vulnerable component is present in the delivered software. However, these later-stage snapshots only indicate that the component exists; they do not explain why it appears in the project. Build-stage dependency resolution further reveals that {jackson-databind@2.13.5} is introduced transitively through {spring-boot-actuator-autoconfigure@2.7.18}, which is itself introduced by {spring-boot-starter-actuator@2.7.18}. Thus, {jackson-databind} is not directly declared by the project, but is introduced indirectly through an upstream Spring Boot dependency chain. For such transitive dependencies, identifying a vulnerable component from a later-stage snapshot alone does not directly reveal an actionable remediation point. Developers must further trace its dependency origin to locate the upstream direct dependency that can actually be upgraded\cite{ponta2020detection,rahkema2023propagation}.

These two paths show that the limitation of {snapshot-based SCA} is not merely incomplete component coverage, but the lack of lifecycle information required to support complete security assessment from any single composition snapshot. Earlier-stage snapshots can reveal where a component is introduced, but cannot determine whether it ultimately reaches the delivered software; later-stage snapshots can confirm its final presence, but often cannot recover the original introduction relationship. Therefore, software composition analysis should answer not only {whether} a component exists, but also two additional questions: {where it comes from} and {where it ultimately ends up} across subsequent software states.

\subsection{Software Composition as Lifecycle Context}

Building on the motivating case above, we define the supply-chain-security-relevant state of a component throughout the software lifecycle as its \emph{component security context}, characterized along two complementary dimensions: Origin and Fate. Origin describes through which dependency relationship or software process a component is introduced into the system, while Fate captures how the component persists across subsequent lifecycle stages and where it ultimately ends up. Compared with a composition snapshot that only records whether a component is present in a particular software state, the security context further connects the component's introduction source with its final state.

Recovering such a security context requires component evidence that supports both Origin and Fate. However, this evidence is not concentrated in any single software object; instead, it emerges progressively as the software transitions from development to delivery and execution. A component's introduction source is typically preserved in dependency declarations, build-resolution results, and dependency relationships \cite{na2024cneps}, whereas its later state must be confirmed through its actual presence in release artifacts, deployment environments, and runtime states ~\cite{xia2023sbom,xiao2025jbomaudit}. Therefore, fully reconstructing a component's security context requires organizing and correlating these distributed composition evidences along the software lifecycle.

Therefore, we first identify the lifecycle states that produce composition evidence with distinct semantics. Rather than directly adopting the complete DevSecOps lifecycle, we abstract it into key states that represent distinguishable software composition states \cite{nist_devsecops_reference}. A typical DevSecOps lifecycle includes stages such as Plan, Code/Develop, Build, Test, Release, Deploy, Operate, and Monitor. The Plan stage does not yet produce analyzable dependency composition; Test typically reuses dependencies resolved and artifacts produced during Build and therefore does not introduce a new primary composition state. Operate and Monitor, meanwhile, jointly reflect the software's actual post-deployment execution state. Based on this abstraction, we summarize the lifecycle states directly relevant to software composition evolution into five stages: {Code}, {Build}, {Release}, {Deploy}, and {Runtime}.

Because these stages correspond to different software states, they expose substantially different forms of component evidence and therefore provide different levels of support for recovering Origin and Fate. This suggests that a \emph{component security context} does not reside in any single lifecycle stage, but must instead be reconstructed from evidence distributed across multiple stages. A single-stage SCA result typically provides only a composition snapshot of one software state: it can indicate whether a component is present, but cannot simultaneously explain its Origin and Fate. Therefore, software composition analysis should move beyond static component snapshots toward a \emph{lifecycle-aware security context} that correlates evidence across lifecycle stages.

\section{Approach}
\label{sec:schema-design}

\subsection{Overview}
\label{sec:design-logic}
\begin{figure*}[t]
    \centering
    \includegraphics[width=0.8\linewidth]{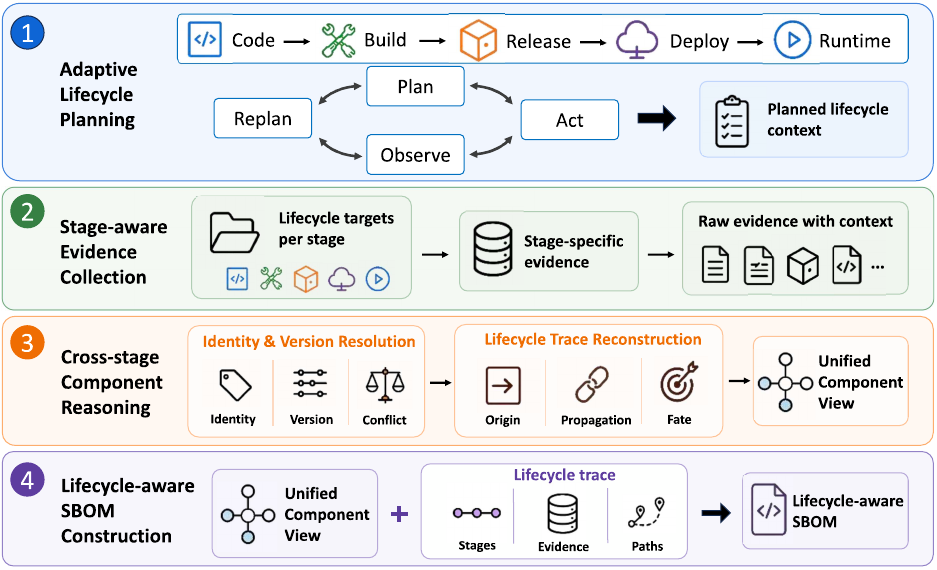}
\caption{Overview of \emph{SCA-Agent.}}   \label{fig:RQ2}
\end{figure*}

To address the limitation of traditional SCA in simultaneously determining a component's \emph{Origin} and \emph{Fate}, \textit{SCA-Agent} dynamically determines which lifecycle stages and analysis targets should be explored based on the target project's structure, build workflow, and analysis feedback. It then collects component-related evidence from different lifecycle stages while preserving the corresponding stage and provenance context. By further correlating component information across stages, \textit{SCA-Agent} reconstructs each component's introduction source, propagation process, and final lifecycle state, and ultimately produces an {SBOM} enriched with lifecycle context.
\textit{SCA-Agent} consists of four interconnected steps:

\noindent $\bullet$ \textbf{Adaptive Lifecycle Planning.}
\textit{SCA-Agent} dynamically plans the lifecycle analysis path based on the target project's structure, build configuration, and analysis feedback, and identifies the software objects at different stages that may contain component information.

\noindent $\bullet$  \textbf{Stage-aware Evidence Collection.}
\textit{SCA-Agent} collects component evidence from different lifecycle stages and preserves contextual information including the stage at which the evidence is produced, its source object, and the associated evidence path.

\noindent $\bullet$  \textbf{Cross-stage Component Reasoning.}
\textit{SCA-Agent} correlates component evidence across stages, resolves component identities and versions, and reconstructs component lifecycle traces from introduction to final state.

\noindent $\bullet$ \textbf{Lifecycle-aware SBOM Construction.}
\textit{SCA-Agent} organizes the reconstructed component lifecycle information into a traceable software composition view, augmenting a traditional SBOM with component lifecycle context.

\subsection{Adaptive Lifecycle Planning}

Adaptive Lifecycle Planning addresses the challenge that a project's lifecycle path is not known in advance. Projects can differ substantially in language ecosystem, build system, release process, and deployment environment, causing component-related evidence to be distributed across different lifecycle stages and software objects. Therefore, \textit{SCA-Agent} does not follow a fixed analysis workflow. Instead, it dynamically explores the lifecycle path based on the target project structure and newly observed information during execution, and determines which stage and object should be analyzed next.
This process consists of two closely connected steps: {Repository Understanding and Initial Planning} and {Adaptive Planning and Execution}.

\noindent \textbf{Repository Understanding and Initial Planning.}
\textit{SCA-Agent} first inspects the target repository to identify engineering information relevant to software composition analysis, including directory structure, source-code types, dependency descriptor files, build configurations, release scripts, and deployment files. Based on this information, \textit{SCA-Agent} infers the project's language ecosystem, package management scheme, build system, and potential software delivery path. For each supported ecosystem, \textit{SCA-Agent}'s base prompt incorporates lifecycle knowledge such as common dependency resolution mechanisms, build workflows, artifact types, and analysis-tool usage strategies. \textit{SCA-Agent} then combines this ecosystem-level prior knowledge with project-specific characteristics to construct an initial lifecycle exploration plan, identifying candidate source dependencies, build outputs, release artifacts, deployment objects, and runtime environments for subsequent analysis. The goal of this step is not to directly recover the final component set, but to establish the project's lifecycle context and locate where component evidence may emerge.

\noindent \textbf{Adaptive Planning and Execution.}
During execution, \textit{SCA-Agent} performs analysis actions according to the current plan and continuously updates its analysis state with the observations produced by each step. When new modules, build outputs, release packages, container images, deployment configurations, or runtime objects are discovered, they are added to the analysis state and used to guide subsequent lifecycle exploration.

When the current analysis path can no longer make progress, or when the current object does not provide sufficient component information, \textit{SCA-Agent} revises its plan based on error logs, project configurations, and the accumulated analysis state. Replanning may involve adjusting execution parameters, selecting alternative tools, switching analysis targets, or exploring newly discovered lifecycle objects. As a result, \textit{SCA-Agent} forms a closed-loop \emph{planning--execution--observation--replanning} process, allowing the analysis workflow to evolve dynamically according to the specific project and previously collected evidence rather than relying on a fixed lifecycle script. To prevent unrecoverable projects from causing unbounded exploration, \textit{SCA-Agent} limits this process to at most 30 iterations.

Finally, Adaptive Lifecycle Planning outputs the identifiable lifecycle stages of the target project, the corresponding analysis objects at each stage, and the derivation relationships among these objects, providing the context required for subsequent stage-aware evidence collection. {Stage-aware Evidence Collection} step.

\subsection{Stage-aware Evidence Collection}

Stage-aware Evidence Collection addresses the challenge that component evidence is distributed across different lifecycle stages. Because components take different observable forms at different stages, evidence from any single stage is insufficient to simultaneously determine a component's Origin and Fate. Therefore, based on the analysis objects identified by Adaptive Lifecycle Planning, \textit{SCA-Agent} collects component evidence with explicit stage semantics across multiple lifecycle stages.

\textit{SCA-Agent}'s base prompt incorporates ecosystem-specific dependency resolution methods, tool usage strategies, and basic analysis rules. According to the current lifecycle stage and analysis target, \textit{SCA-Agent} dynamically selects the corresponding analysis operations to extract component-related information from Code, Build, Release, Deploy, and Runtime.
During execution, \textit{SCA-Agent} selects and adapts parsing scripts for the target object to obtain observable information such as component identity, version, and dependency relationships. For example, the Code stage exposes dependency declarations, the Build stage provides resolved dependency information, while the Release, Deploy, and Runtime stages further reveal whether components reach the final software environment.
Because component representations and evidence granularity differ across stages, \textit{SCA-Agent} does not directly merge cross-stage results during evidence collection. Instead, it preserves each raw observation together with contextual metadata, including the {source stage}, {evidence source}, and {evidence path}.

Finally, the outputs from each stage are organized as {stage-specific evidence}. This representation preserves the local state of a component at a particular lifecycle stage while enabling each analysis result to be traced back to the corresponding file, artifact, or runtime object, providing the foundation for subsequent cross-stage reasoning.

\subsection{Cross-stage Component Reasoning}

Cross-stage Component Reasoning is the core module in \textit{SCA-Agent} for recovering component lifecycle context. Its goal is to associate local component observations across different lifecycle stages and reconstruct the complete trajectory of a component from introduction to final state.
Because component evidence from different stages may differ in naming conventions, version representations, and dependency context, \textit{SCA-Agent} divides this process into two steps: {Identity and Version Resolution} and {Lifecycle Trace Reconstruction}.

\noindent \textbf{Identity and Version Resolution.}
This step first determines whether component observations from different lifecycle stages refer to the same software entity. Specifically, \textit{identity normalization} maps package coordinates, file names, artifact metadata, and runtime records from different stages to a unified component identity. Through this process, \textit{SCA-Agent} reconciles heterogeneous component representations across stages into unified software component entities.

It then determines which concrete version of each component is ultimately used. \textit{Version resolution} combines source-level declarations, dependency resolution results, and downstream artifact information to resolve variables, inheritance relationships, and version ranges into concrete component versions. \textit{Dependency conflict resolution} further handles cases in which the same component is introduced through multiple dependency paths with different candidate versions. \textit{SCA-Agent} uses the build system's actual resolution result together with downstream lifecycle evidence to determine the version that is ultimately selected and delivered, while preserving its original declaration and dependency source.

\noindent \textbf{Lifecycle Trace Reconstruction.}
After resolving component identities and versions, \textit{SCA-Agent} reconstructs how each component propagates through the software lifecycle.
It first determines how the component is introduced based on stage-specific evidence, including direct declarations in source code, indirect introduction through dependencies, or components that first appear at later lifecycle stages. By analyzing the stage in which a component first appears together with its dependency context, \textit{SCA-Agent} recovers its \emph{Origin}.

\textit{SCA-Agent} then correlates component evidence across lifecycle stages and combines this evidence with derivation relationships among analysis objects to reconstruct cross-stage propagation, determining how a component moves from source code and the build environment into release artifacts, deployment environments, and runtime objects. Finally, it uses evidence from subsequent stages to infer the component's Fate, including whether it is removed during later processing, included in the final delivered artifact, deployed, or loaded or executed at Runtime.
Ultimately, Cross-stage Component Reasoning transforms component observations from lifecycle stages into evidence-backed lifecycle traces, which serve as the input to Lifecycle-aware SBOM Construction.

\subsection{Lifecycle-aware SBOM Construction}

Lifecycle-aware SBOM Construction organizes the unified component view produced by {Cross-stage Component Reasoning} into the final Lifecycle-aware SBOM. \textit{SCA-Agent} first generates fundamental software composition information, including component identity, version, and dependency relationships, following standard SBOM formats. On top of these standard fields, each dependency entry is further augmented with its recovered complete lifecycle trajectory.

Specifically, in addition to standard SBOM fields, each dependency entry records its lifecycle states across Code, Build, Release, Deploy, and Runtime stages, the source of dependency introduction, and the propagation relationships among different stages. Furthermore, each entry is associated with the corresponding {stage-specific evidence} and evidence paths that support these lifecycle decisions.

Finally, \textit{SCA-Agent} outputs a unified Lifecycle-aware SBOM that extends traditional static component inventories with additional information about where a component is introduced, which lifecycle stages it passes through, and what final state it reaches. This enhanced representation provides a foundation for downstream component auditing, vulnerability analysis, and software supply chain security analysis.
\section{Experiment}
\label{sec:Experiment}

Our experiments focus on answering the following \emph{Research Questions} (RQs):






\begin{itemize}[
  leftmargin=*,
  labelsep=0.5em,
  itemsep=0.3em,
  topsep=0.3em,
  parsep=0pt
]
  \item \textbf{RQ1:} How do software dependencies evolve across lifecycle stages?

  \item \textbf{RQ2:} How necessary are the components of \emph{SCA-Agent}?

  \item \textbf{RQ3:} How effectively does \emph{SCA-Agent} identify software components across lifecycle stages?

  \item \textbf{RQ4:} How accurately does \emph{SCA-Agent} recover dependency component lifecycle trajectories?

  \item \textbf{RQ5:} Does lifecycle context improve software supply-chain vulnerability exposure assessment?

\end{itemize}
\subsection{Experimental Setup}
\textbf{Lifecycle-aware Ground Truth Construction}. 
We collect open-source projects from GitHub across three ecosystems: Java, JavaScript, and Python. We require each project to have at least 100 GitHub stars to ensure a reasonable level of maturity and practical adoption. We then manually validate the collected projects and exclude those for which complete lifecycle analysis cannot be performed due to build failures, missing release artifacts, abnormal dependency management, or toolchain incompatibility. Finally, we retain 35 projects for each language ecosystem, resulting in a total of 105 experimental subjects. The selected projects cover a variety of software categories, including Web frameworks, build tools, enterprise applications, microservice systems, and data science libraries, and exhibit substantial diversity in project size, dependency scale, and build complexity.

For each project, we perform a complete analysis across the five lifecycle stages defined in this work, namely Code, Build, Release, Deploy, and Runtime, and collect the third-party dependencies actually present at each stage. Specifically, at the Code stage, we comprehensively collect and parse dependency declaration files in the project whenever possible. At the Build stage, we record the dependency components actually resolved and downloaded during the build process. At the Release stage, we analyze the generated release artifacts. At the Deploy stage, we identify relevant components in the actual deployment environment by combining information from Dockerfiles, deployment configurations, and project documentation. At the Runtime stage, we manually construct inputs that cover as much project functionality as possible, execute the project, and identify dependencies that are actually loaded or used through runtime monitoring. In total, we obtain 58,211 stage-level dependency instances.

Based on the above process, we first organize the dependencies collected at each lifecycle stage into the {stage-level component ground truth}, which characterizes the actual software composition at different lifecycle stages and serves as the reference for subsequent stage-level component detection and lifecycle evolution analysis. Second, we further correlate the presence states, dependency origins, and cross-stage propagation relationships of the same component across different stages to construct the lifecycle trace ground truth. This ground truth records the introduction origin and final lifecycle state of each component and is used to evaluate the accuracy of recovering component Origin and Fate. Finally, we associate the components confirmed at each stage with their corresponding vulnerability information to construct the CVE-stage ground truth. Each $(\mathrm{CVE}, \mathrm{stage})$ pair is treated as a basic annotation unit, indicating the actual presence and exposure state of a vulnerability at a specific lifecycle stage, and is used to evaluate the impact of lifecycle context on software supply chain vulnerability detection. In total, we construct 17,482 CVE-stage ground-truth instances.

To ensure ground-truth reliability, three annotators independently extract and review dependencies at each lifecycle stage. Disagreements are resolved through discussion and consensus. The resulting annotations form the stage-level component sets, lifecycle-level component sets, and component lifecycle trace ground truth.

\noindent\textbf{Baseline Configuration.}
To evaluate \emph{SCA-Agent} against existing SCA approaches, we select {Syft \cite{anchore_syft}, Microsoft SBOM Tool \cite{microsoft_sbom_tool}, and cdxgen \cite{cyclonedx_cdxgen}} as baselines. All three are representative open-source SCA/SBOM tools but adopt different component discovery strategies. {Syft} primarily scans software artifacts for components, {Microsoft SBOM Tool} detects components from project files and build-related information, and {cdxgen} recovers direct and transitive dependencies mainly from manifests, lockfiles, and package-manager information. Together, they represent artifact scanning, project dependency detection, and package-manager-driven component recovery, providing representative baselines for comparing different SCA strategies.

\noindent\textbf{\emph{SCA-Agent} Implementation.}
\emph{SCA-Agent} is implemented based on the {Claude Agent framework} \cite{anthropicClaudeAgent} and organizes its lifecycle analysis capabilities in a {skill-driven} manner. Specifically, we design a dedicated skill for each of the four core analysis steps described in our approach, with corresponding domain knowledge, processing strategies, and auxiliary scripts configured in their references. This design enables the agent to perform lifecycle planning, stage-aware analysis, cross-stage reasoning, and final result construction following the defined analysis workflow. In our experiments, we use \emph{Claude-Sonnet-4.6} \cite{anthropicSonnet46} as the underlying language model and maintain the same model configuration across all experiments. \emph{SCA-Agent} ultimately produces a {Lifecycle-aware SBOM} conforming to the CycloneDX standard, recording both component information and its lifecycle context to support subsequent lifecycle reconstruction and security analysis.

\noindent\textbf{Evaluation Metrics.}
To evaluate the lifecycle analysis capability of \emph{SCA-Agent}, we define task-specific evaluation metrics for different analysis objectives. For lifecycle component evolution, we use {Added}, {Removed}, {Retained}, and Jaccard Similarity to characterize component changes and set overlap between adjacent lifecycle stages. For stage-level component identification and mechanism ablation experiments, predicted components are matched against the ground truth of the corresponding stage, and project-level average Precision, Recall, and F1-score are calculated based on TP, FP, and FN. For lifecycle trajectory reconstruction, we use Origin Accuracy and Fate Accuracy to measure the correctness of identifying the earliest stage in which a dependency appears and the final stage in which it is retained, respectively. For vulnerability exposure detection, each exposure instance is defined by a component, its associated vulnerability, and the lifecycle stage in which the vulnerability is present; Precision, Recall, and F1-score are then calculated based on TP, FP, and FN.

\subsection{Lifecycle Dependency Evolution}

To analyze how software dependencies evolve throughout the lifecycle and whether a single lifecycle stage can represent the complete software composition, we use the {stage-level component ground truth} to examine component evolution across the five stages of Code, Build, Release, Deploy, and Runtime for Java, JavaScript, and Python projects. Specifically, we measure the numbers of {Added}, {Removed}, and {Retained} components between adjacent stages to characterize component introduction, removal, and persistence throughout the lifecycle. We further compute {Jaccard Similarity} to quantify the overlap between component sets across different stages.


\begin{table}[t]
\centering
\caption{Distribution of component observations across lifecycle stages.}
\label{tab:stage_component_count}

\setlength{\tabcolsep}{5pt}
\renewcommand{\arraystretch}{1.15}

\resizebox{\columnwidth}{!}{
\begin{tabular}{@{}ccccccc@{}}
\toprule
Language & Code & Build & Release & Deploy & Runtime & Total \\
\midrule
Python     & 993  & 646   & 748  & 678  & 388  & 3453  \\
JavaScript & 1145 & 21784 & 370  & 1075 & 726  & 25100 \\
Java       & 1753 & 10285 & 6945 & 7056 & 3619 & 29658 \\
\midrule
Total      & 3891 & 32715 & 8063  & 8809 & 4733 & 58211 \\
\bottomrule
\end{tabular}
}
\end{table}
{Software dependencies are continuously introduced, removed, and adjusted as the lifecycle progresses, rather than remaining static.} As shown in Table \ref{tab:stage_component_count}, we obtain a total of {58,211 stage-level component observations} across the five lifecycle stages. Among them, the Build stage contains the largest number of components ({32,715}), substantially exceeding the Code stage ({3,891}), indicating that the build process introduces a large number of dependencies that are not visible at the source-code stage. From Build to Release, the number of components decreases to {8,063}, showing that many build-time dependencies do not enter the final release artifacts. The number then increases again to {8,809} at the Deploy stage, while further decreasing to {4,733} at Runtime, indicating that deployed components are not necessarily equivalent to runtime components. The Added and Removed results further show that components continuously transition throughout the lifecycle, and their security states cannot be fully captured from any single stage.

{Component sets are not consistently stable across adjacent lifecycle stages, indicating that no single stage can fully represent the overall software composition.} As shown in Table~2, the number of {Retained} components in each stage transition is consistently lower than the complete component set of the corresponding stages, indicating that only a subset of dependencies persists into the next lifecycle stage, while the remaining components are removed or replaced. Meanwhile, Jaccard Similarity further shows that the overall compositions of adjacent stages are not consistent. For example, the Jaccard similarities across lifecycle transitions in Python range only from {22.57\%--48.10\%}. Even though Java reaches a relatively high similarity of {98.29\%} from Release to Deploy, this does not imply comparable stability across the remaining stages. Taken together, the Retained and Jaccard results show that component sets are continuously reorganized as the lifecycle progresses, and any individual stage can only reflect a partial composition state rather than fully represent the complete software composition.








\begin{table}[htbp]
\centering
\caption{Dependency changes between different software lifecycle stages.
C, B, Rel, D, and Run denote Code, Build, Release, Deploy, and Runtime stages, respectively.}
\label{tab:dependency_changes}

\small
\renewcommand{\arraystretch}{1.15}
\setlength{\tabcolsep}{2pt}

\begin{tabularx}{\columnwidth}{
  @{}
  >{\centering\arraybackslash}p{0.16\columnwidth}
  >{\centering\arraybackslash}p{0.13\columnwidth}
  *{4}{>{\centering\arraybackslash}X}
  @{}
}
\toprule
\multirow{2}{*}{Language}
& \multirow{2}{*}{Measure}
& \multicolumn{4}{c}{Lifecycle Transition} \\
\cmidrule(lr){3-6}
&
& C$\rightarrow$B
& B$\rightarrow$Rel
& Rel$\rightarrow$D
& D$\rightarrow$Run \\
\midrule

\multirow{4}{*}{JavaScript}
& Added    & 506.74 & 0.03   & 18.89 & 0.29 \\
& Removed  & 1.26   & 527.49 & 0.03  & 9.63 \\
& Retained & 31.29  & 10.54  & 10.54 & 19.80 \\
& Jaccard  & 8.10\% & 2.44\% & 53.07\% & 84.68\% \\
\cmidrule(lr){1-6}

\multirow{4}{*}{Java}
& Added    & 212.94 & 2.89   & 3.17   & 1.00 \\
& Removed  & 1.23   & 65.83  & 0      & 97.77 \\
& Retained & 46.49  & 193.60 & 196.49 & 101.89 \\
& Jaccard  & 18.10\% & 73.38\% & 98.29\% & 55.30\% \\
\cmidrule(lr){1-6}

\multirow{4}{*}{Python}
& Added    & 9.26  & 10.49 & 11.17 & 1.43 \\
& Removed  & 15.71 & 9.00  & 11.74 & 9.71 \\
& Retained & 9.20  & 9.46  & 8.20  & 9.66 \\
& Jaccard  & 26.71\% & 29.78\% & 22.57\% & 48.10\% \\

\bottomrule
\end{tabularx}
\end{table}

\begin{tcolorbox}[
    colback=gray!5,
    colframe=gray!45,
    boxrule=0.5pt,
    arc=1.5pt,
    left=5pt,
    right=5pt,
    top=4pt,
    bottom=4pt,
    title=\textbf{Summary},
    coltitle=black,
    colbacktitle=gray!15,
    fonttitle=\bfseries,
    enhanced,
    attach boxed title to top left={xshift=4pt,yshift=-2pt},
    boxed title style={
        colback=gray!15,
        colframe=gray!45,
        boxrule=0.4pt,
        arc=1.5pt
    }
]
Software dependencies are continuously introduced, retained, and removed as the lifecycle progresses, while both the number of Retained components and the overall set similarity between adjacent stages indicate that software composition remains unstable across stages. Therefore, no single lifecycle stage can fully represent the actual state of components throughout the entire lifecycle.
\end{tcolorbox}

\subsection{Ablation Study}

To analyze the contributions of the key mechanisms in \emph{SCA-Agent} to lifecycle component recovery, we further conduct an ablation study. Specifically, we remove the two core modules, {Adaptive Lifecycle Planning} and {Cross-stage Component Reasoning}, to construct two variants, {No Planning} and {No Reasoning}, respectively, and evaluate them on the same projects under the same execution environment. All variants are evaluated against the {stage-level component ground truth} using Precision, Recall, and F1 to assess how different mechanisms affect component recovery across lifecycle stages.

{No Planning} removes the dynamic planning process based on project structure and previous analysis results, and instead uses a fixed prompt that instructs the agent to analyze the Code, Build, Release, Deploy, and Runtime stages sequentially. This variant is used to evaluate the impact of adaptive analysis paths on lifecycle tracking. 

{No Reasoning} preserves the independent analysis of each lifecycle stage but removes cross-stage component association and evidence integration, generating the final component view solely from stage-level results. This variant is used to evaluate the contribution of cross-stage reasoning to component identity resolution and lifecycle recovery.

{Adaptive planning enables continuous lifecycle evidence discovery.} As shown in Fig.~\ref{fig:RQ2}, removing Planning causes substantial performance degradation across all three language ecosystems, with the most pronounced impact occurring in later lifecycle stages such as Release, Deploy, and Runtime.

For Java, No Planning decreases F1 by {52.84, 85.87, 88.66, 86.56, and 77.80 percentage points} across the five stages, respectively, with post-build stages being affected most severely. For JavaScript, performance remains relatively high at the Code and Build stages, whereas F1 at the Release and Deploy stages decreases by {89.40 and 85.04 percentage points}, respectively. Python exhibits a similar trend after the Build stage, with F1 at the Release and Deploy stages decreasing by {78.48 and 67.05 percentage points}, respectively.

These results show that although a fixed prompt can leverage explicit dependency declarations or build configurations to identify components in some early stages, it cannot dynamically adjust subsequent analysis targets according to project structure and previously obtained analysis results. Since component evidence in later lifecycle stages is typically distributed across heterogeneous objects such as release artifacts, deployment configurations, and runtime environments, the absence of Planning limits the agent's ability to continuously locate downstream analysis objects, preventing the lifecycle evidence chain from being fully extended.

\begin{figure*}[t]
    \centering
    \includegraphics[width=1\linewidth]{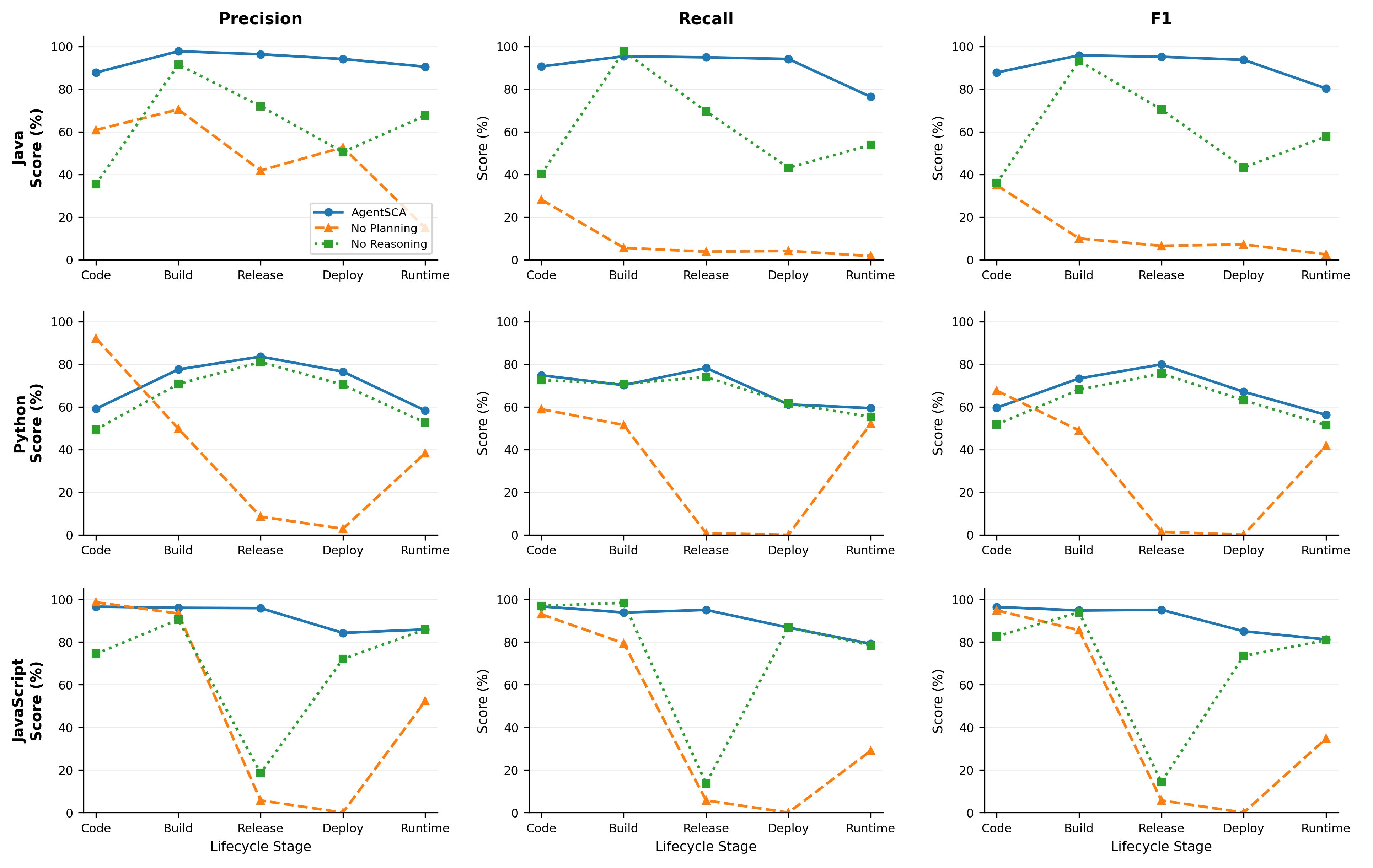}
\caption{Ablation Results Across Lifecycle Stages and Language Ecosystems}   \label{fig:RQ2}
\end{figure*}

{Cross-stage reasoning enables component identity resolution and lifecycle association.} After removing Cross-stage Reasoning, the performance degradation mainly reflects weakened component identity resolution and cross-stage association capabilities. As shown in Fig.~\ref{fig:RQ2}, No Reasoning exhibits decreases in both Precision and Recall across multiple lifecycle stages, indicating that independently obtained stage-level results cannot be directly transformed into an accurate lifecycle component view.

For component identity resolution, the lack of cross-stage evidence constraints makes it difficult to associate components with similar names but different versions or origins across stages. For example, Precision at the Java Code stage decreases by {52.39 percentage points}, while Precision at the JavaScript Release stage decreases by {77.38 percentage points}. This indicates that relying on single-stage detection results can easily introduce incorrect matches, whereas cross-stage evidence provides constraints such as component versions, dependency relationships, and occurrence locations.

For continuous lifecycle tracking, removing cross-stage reasoning also weakens component recovery in later stages. Recall at the Java Deploy and Runtime stages decreases by {50.96 and 22.60 percentage points}, respectively, while Recall at the JavaScript Release stage decreases by {81.37 percentage points}. This is because components in later stages often need to be completed and validated using dependency relationships established in earlier stages, whereas simply aggregating independent stage-level results cannot determine the evolutionary relationships among components.

Planning determines whether \emph{SCA-Agent} can continuously discover component evidence across lifecycle stages, while Cross-stage Reasoning determines whether such distributed evidence can be correctly associated and reconstructed into complete lifecycle traces. Together, these two mechanisms enable \emph{SCA-Agent} to move beyond stage-level component detection toward lifecycle-aware component recovery. 

\begin{tcolorbox}[
    colback=gray!5,
    colframe=gray!45,
    boxrule=0.5pt,
    arc=1.5pt,
    left=5pt,
    right=5pt,
    top=4pt,
    bottom=4pt,
    title=\textbf{Summary},
    coltitle=black,
    colbacktitle=gray!15,
    fonttitle=\bfseries,
    enhanced,
    attach boxed title to top left={xshift=4pt,yshift=-2pt},
    boxed title style={
        colback=gray!15,
        colframe=gray!45,
        boxrule=0.4pt,
        arc=1.5pt
    }
]
Planning enables \emph{SCA-Agent} to continuously discover lifecycle-specific evidence, while cross-stage reasoning integrates heterogeneous evidence to resolve component identity and reconstruct complete lifecycle contexts.
\end{tcolorbox}

\subsection{Component Detection Comparison}

To evaluate the component detection capability of \emph{SCA-Agent} across different lifecycle stages, we compare \emph{SCA-Agent} with traditional SCA tools using the {stage-level component ground truth}. For each of the five stages, namely {Code}, {Build}, {Release}, {Deploy}, and {Runtime}, we match the component set reported by each tool against the corresponding stage-level ground truth and use the average F1 score to measure detection accuracy. Since traditional SCA tools typically generate component inventories from specific software objects without explicitly modeling lifecycle stages, we align their detection results with the ground truth of each of the five stages to assess their component coverage across different lifecycle stages.

\begin{figure*}[htbp]
    \centering
    \begin{minipage}{0.33\textwidth}
        \centering
        \includegraphics[width=\linewidth]{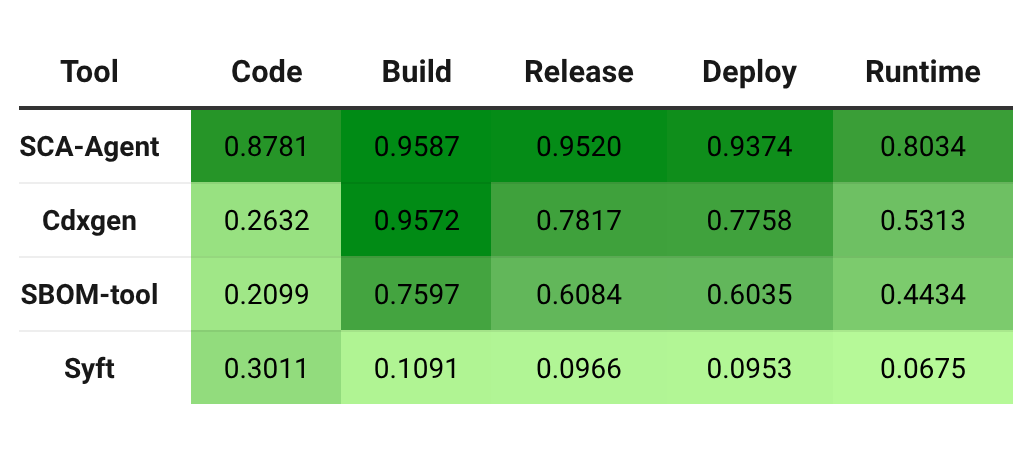}
        \par\small (a) Java
    \end{minipage}
    \hfill
    \begin{minipage}{0.33\textwidth}
        \centering
        \includegraphics[width=\linewidth]{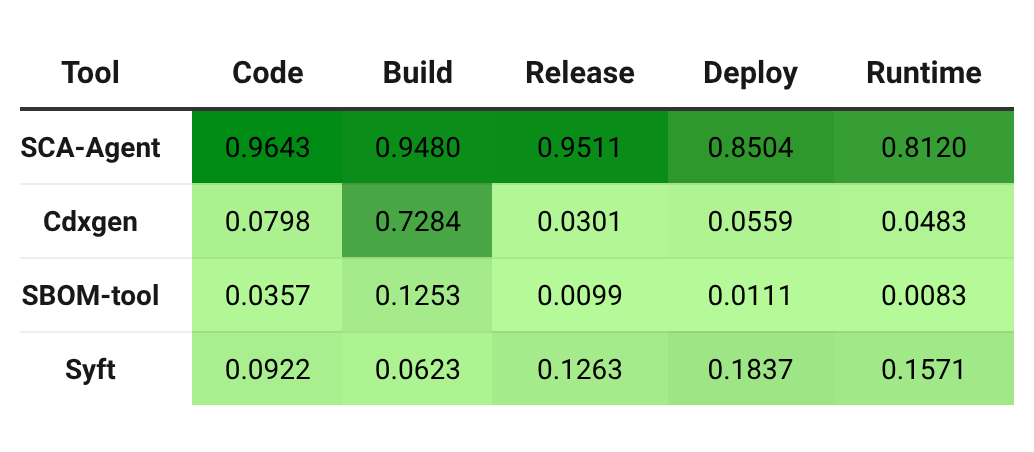}
        \par\small (b) JavaScript
    \end{minipage}
    \hfill
    \begin{minipage}{0.33\textwidth}
        \centering
        \includegraphics[width=\linewidth]{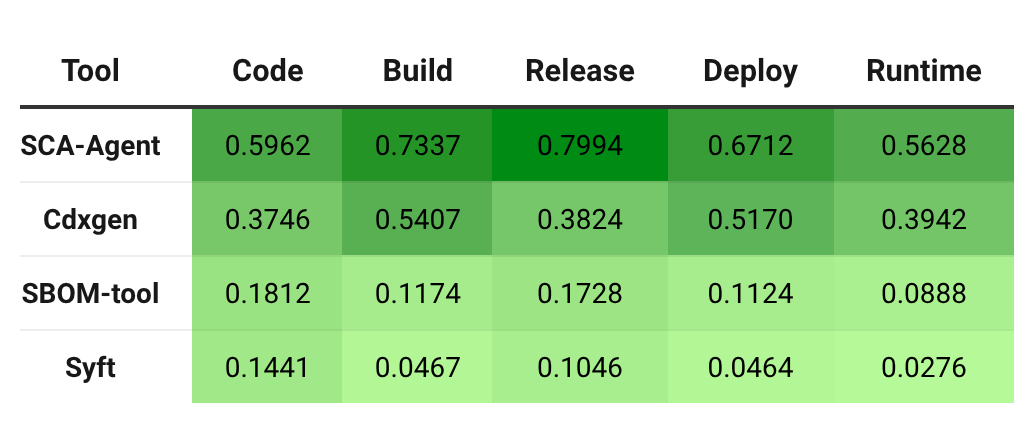}
        \par\small (c) Python
    \end{minipage}
    
    \caption{Stage-level component identification performance of SCA tools across language ecosystems.}
    \label{fig:heatmaps}
\end{figure*}

To evaluate the component detection capability of \emph{SCA-Agent} across different lifecycle stages, we compare \emph{SCA-Agent} with Syft, SBOM-tool, and Cdxgen based on the {stage-level component ground truth}. For the five lifecycle stages, including Code, Build, Release, Deploy, and Runtime, we match the component sets reported by each tool at the corresponding stage with the ground truth of that stage, and use the average F1 score to measure detection accuracy. Since traditional SCA tools typically generate component inventories based on specific software artifacts without explicitly modeling lifecycle stages, we align their detection results with the ground truth of each lifecycle stage to analyze their component coverage capability throughout the software lifecycle.

As shown in Fig.~\ref{fig:heatmaps}, \emph{SCA-Agent} achieves the highest F1 across all five lifecycle stages, namely {Code}, {Build}, {Release}, {Deploy}, and {Runtime}, while maintaining relatively stable detection performance from early to later stages. In contrast, traditional SCA tools perform mainly well at the Code and Build stages, but generally degrade substantially at the Release, Deploy, and Runtime stages. For example, cdxgen achieves an F1 of {0.957} at the Java Build stage, which is close to \emph{SCA-Agent}'s {0.959}, but its performance drops considerably at the subsequent Release, Deploy, and Runtime stages. This is because traditional tools typically recover component sets from dependency manifests, package-manager metadata, or build systems at a particular stage, without continuously tracking how dependencies evolve afterward. {In contrast, \emph{SCA-Agent} analyzes stage-specific project files, configurations, and runtime environment information across different lifecycle stages, allowing it to maintain high component detection performance even in later stages.}

{Across different language ecosystems, traditional SCA tools exhibit more pronounced language-specific performance variations, whereas \emph{SCA-Agent} maintains a relatively consistent advantage across Java, Python, and JavaScript.} The Java ecosystem provides relatively standardized dependency resolution and build mechanisms through Maven and Gradle, enabling traditional tools to achieve strong results at certain stages; for example, cdxgen already approaches \emph{SCA-Agent} at the Java Build stage. By comparison, dependency sources and project structures in Python and JavaScript are more diverse and may be jointly affected by dependency files, development dependencies, optional dependencies, build scripts, and environment configurations, making fixed analysis workflows difficult to apply consistently across projects. {In contrast, \emph{SCA-Agent} dynamically selects analysis objects and execution strategies according to the specific language ecosystem and project structure, enabling it to achieve the highest F1 in all three languages while exhibiting smaller cross-language performance variations.}

\begin{tcolorbox}[
    colback=gray!5,
    colframe=gray!45,
    boxrule=0.5pt,
    arc=1.5pt,
    left=5pt,
    right=5pt,
    top=4pt,
    bottom=4pt,
    title=\textbf{Summary},
    coltitle=black,
    colbacktitle=gray!15,
    fonttitle=\bfseries,
    enhanced,
    attach boxed title to top left={xshift=4pt,yshift=-2pt},
    boxed title style={
        colback=gray!15,
        colframe=gray!45,
        boxrule=0.4pt,
        arc=1.5pt
    }
]
\emph{SCA-Agent} achieves the highest F1 across all lifecycle stages and all three language ecosystems, demonstrating both stable component detection performance across different stages and strong adaptability to variations in dependency management and build processes across language ecosystems.
\end{tcolorbox}


\subsection{Lifecycle Trajectory Recovery}
\label{tab:RQ4}
This experiment evaluates \emph{SCA-Agent}'s capability to recover dependency {Origin} and {Fate} using the manually constructed {lifecycle trace ground truth}. For traditional SCA tools, we derive their inferred lifecycles from the stage-level detection results in RQ3: for each language ecosystem, we treat the stage with the highest F1 as the latest stage at which a component can be confirmed to exist, and assume that the component is also present in all preceding stages. We then compare \emph{SCA-Agent} with Syft, Microsoft SBOM Tool, and cdxgen across Java, JavaScript, and Python, using {Origin Accuracy} and {Fate Accuracy} to measure the accuracy of identifying the starting and ending points of component lifecycles.

\textbf{For Origin estimation.} As show in Table \ref{tab:origin_fate} that {\emph{SCA-Agent} substantially improves Origin localization accuracy across all three language ecosystems.} In Python, JavaScript, and Java, \emph{SCA-Agent} achieves Origin Accuracy values of {84.44\%, 99.11\%, and 90.65\%}, respectively, significantly outperforming all traditional tools. In comparison, Syft achieves its best result of only {7.05\%} on Java, Microsoft SBOM Tool reaches {28.78\%} on Python, and cdxgen reaches {43.62\%} on Python. This advantage mainly comes from \emph{SCA-Agent}'s cross-stage component reasoning mechanism. For components observed in later stages, \emph{SCA-Agent} can correlate dependency declarations, resolution relationships, and propagation evidence from preceding stages, unify component identities across stages, and trace them back to their original introduction points, thereby recovering component Origin more accurately.

\begin{table}[htbp]
\centering
\caption{Origin and Fate scores by language and tool.}
\label{tab:origin_fate}

\setlength{\tabcolsep}{11pt}
\renewcommand{\arraystretch}{1.15}

\begin{tabular}{c c r r}
\toprule
\multirow{2}{*}{Language} 
    & \multirow{2}{*}{Tool} 
    & \multicolumn{2}{c}{Accuracy  (\%)} \\
\cmidrule(lr){3-4}
    & & Origin & Fate \\
\midrule

\multirow{4}{*}{JavaScript}
    & SCA-Agent  & 99.11 & 97.95 \\
    & Cdxgen     &  5.99 & 71.55 \\
    & Syft       &  1.56 &  0.66 \\
    & SBOM Tool  &  2.92 & 12.59 \\
\cmidrule(lr){1-4}

\multirow{4}{*}{Java}
    & SCA-Agent  & 90.65 & 70.94 \\
    & Cdxgen     & 16.72 & 26.37 \\
    & Syft       &  7.05 &  1.85 \\
    & SBOM Tool  &  9.03 & 11.57 \\
\cmidrule(lr){1-4}

\multirow{4}{*}{Python}
    & SCA-Agent  & 84.44 & 74.32 \\
    & Cdxgen     & 43.62 &  0.00 \\
    & Syft       &  0.00 &  0.00 \\
    & SBOM Tool  & 28.78 &  3.48 \\

\bottomrule
\end{tabular}
\end{table}

{For Fate estimation}, \emph{SCA-Agent} achieves Fate Accuracy values of {74.32\%, 97.95\%, and 70.94\%} on Python, JavaScript, and Java, respectively. In particular, it reaches {97.95\%} on JavaScript, substantially improves over the traditional tools' results of below {3.5\%} on Python, and also clearly outperforms cdxgen's {26.37\%} on Java. This advantage is mainly attributable to \emph{SCA-Agent}'s adaptive lifecycle planning mechanism. Traditional tools are typically limited to specific software objects or lifecycle stages, whereas \emph{SCA-Agent} can dynamically plan and continuously extend evidence collection into subsequent stages based on the current analysis results. This allows it to determine whether components continue to propagate into Release, Deploy, and Runtime, thereby recovering their final lifecycle states more accurately.

In contrast, traditional SCA tools exhibit clear limitations in recovering component Origin and Fate. Their results are typically derived from specific lifecycle stages or software objects, making it difficult to establish continuous relationships for components across different stages. In particular, for Python, Syft achieves an Origin Accuracy of only 0.00\%, while both cdxgen and Syft obtain 0.00\% Fate Accuracy, and Microsoft SBOM Tool reaches only 3.48\%. These results indicate that even when traditional tools can detect certain components, they often cannot further determine where those components were initially introduced or whether they eventually propagate to later lifecycle stages. In other words, without cross-stage correlation, traditional SCA is better suited to producing local component snapshots than reconstructing complete component lifecycle trajectories.

\begin{tcolorbox}[
    colback=gray!5,
    colframe=gray!45,
    boxrule=0.5pt,
    arc=1.5pt,
    left=5pt,
    right=5pt,
    top=4pt,
    bottom=4pt,
    title=\textbf{Summary},
    coltitle=black,
    colbacktitle=gray!15,
    fonttitle=\bfseries,
    enhanced,
    attach boxed title to top left={xshift=4pt,yshift=-2pt},
    boxed title style={
        colback=gray!15,
        colframe=gray!45,
        boxrule=0.4pt,
        arc=1.5pt
    }
]
\emph{SCA-Agent} achieves {84.44\%--99.11\% Origin Accuracy} and {70.94\%--97.95\% Fate Accuracy} across the three language ecosystems, substantially outperforming traditional SCA tools and demonstrating its ability to more accurately recover dependency origins and propagation states.
\end{tcolorbox}

\subsection{Lifecycle-aware Vulnerability Assessment}
This experiment evaluates whether lifecycle context can improve software supply chain vulnerability exposure assessment. Based on the stage-level vulnerability ground truth defined in the setup, we compare the vulnerability detection results of \emph{SCA-Agent} with those of cdxgen, Microsoft SBOM Tool, and Syft. For traditional tools, the lifecycle stages associated with their detected CVEs follow the component-to-stage mappings established in Section \ref{tab:RQ4}, ensuring consistent lifecycle determination across experiments. We use {Precision}, {Recall}, and {F1} to measure each method's ability to determine the actual exposure state of vulnerabilities.

As shown in the Table \ref{tab:performance}, {\emph{SCA-Agent} achieves the best overall performance in vulnerability exposure assessment.} Among the {17,482} ground-truth vulnerability exposure instances, \emph{SCA-Agent} correctly identifies {16,497}, with only {145 false positives (FPs)} and {985 false negatives (FNs)}. Its Precision, Recall, and F1 reach {99.13\%, 94.37\%, and 96.69\%}, respectively. In comparison, although cdxgen achieves a Recall of {99.30\%}, it produces {9,710 FPs}, resulting in a Precision of only {64.13\%} and an F1 of {77.93\%}. \emph{SCA-Agent} therefore improves F1 by {18.76 percentage points} over the best-performing traditional tool.

\begin{table}[htbp]
\centering
\caption{Performance metrics of different tools}
\label{tab:performance}

\renewcommand{\arraystretch}{1.15}
\setlength{\tabcolsep}{6pt}

\resizebox{\columnwidth}{!}{%
\begin{tabular}{l c c c c}
\toprule
Metric & SCA-Agent & Cdxgen & SBOM Tool & Syft \\
\midrule
Results   & 16,642 & 27,070 & 18,415 & 2,022 \\
TP        & 16,497 & 17,360 & 12,235 & 1,560 \\
FP        & 145    & 9,710  & 6,180  & 462 \\
FN        & 985    & 122    & 5,247  & 15,922 \\
\midrule
Precision & 99.13\% & 64.13\% & 66.44\% & 77.15\% \\
Recall    & 94.37\% & 99.30\% & 69.99\% & 8.92\% \\
F1        & 96.69\% & 77.93\% & 68.17\% & 16.00\% \\
\bottomrule
\end{tabular}%
}
\end{table}

This advantage is directly related to \emph{SCA-Agent}'s accurate recovery of {Origin} and {Fate} in RQ4.1. Accurate Origin allows \emph{SCA-Agent} to determine where a vulnerable component enters the software lifecycle, while accurate Fate enables it to determine whether the component continues to propagate into subsequent stages. As a result, \emph{SCA-Agent} can avoid treating vulnerabilities that have already disappeared during build, release, or deployment as still exposed in later stages, while also reducing omissions caused by incorrect lifecycle boundary estimation. The high Recall but low Precision of cdxgen further shows that extending vulnerability states from traditional static SBOM results can cover most real vulnerabilities but tends to cause substantial over-propagation. In contrast, by relying on reconstructed lifecycle boundaries, \emph{SCA-Agent} reduces the number of FPs from {9,710} to {145}.

Microsoft SBOM Tool achieves Precision, Recall, and F1 values of {66.44\%, 69.99\%, and 68.17\%}, respectively, while still producing a considerable number of both FPs and FNs. Syft achieves only {8.92\%} Recall, resulting in an F1 of {16.00\%}. These results indicate that, without reliable lifecycle context, traditional SCA tools struggle to determine whether vulnerabilities truly persist into subsequent software states. In contrast, \emph{SCA-Agent} combines dependency origin, propagation, and final lifecycle state to more accurately characterize the actual scope of vulnerability exposure.

\begin{tcolorbox}[
    colback=gray!5,
    colframe=gray!45,
    boxrule=0.5pt,
    arc=1.5pt,
    left=5pt,
    right=5pt,
    top=4pt,
    bottom=4pt,
    title=\textbf{Summary},
    coltitle=black,
    colbacktitle=gray!15,
    fonttitle=\bfseries,
    enhanced,
    attach boxed title to top left={xshift=4pt,yshift=-2pt},
    boxed title style={
        colback=gray!15,
        colframe=gray!45,
        boxrule=0.4pt,
        arc=1.5pt
    }
]
Lifecycle context can substantially improve software supply chain vulnerability exposure assessment. By accurately recovering dependency {Origin} and {Fate}, \emph{SCA-Agent} more precisely identifies the actual propagation and persistence of vulnerabilities throughout the software lifecycle, reducing both the erroneous extension of vulnerabilities that have already disappeared and omissions of real exposures. As a result, \emph{SCA-Agent} achieves an F1 of {96.69\%}, outperforming traditional SCA tools by {18.76--80.69 percentage points}.
\end{tcolorbox}
\section{Discussion}

\noindent \textbf{Toward Self-Evolving \textit{SCA-Agent}.}
\textit{SCA-Agent} currently relies on manually constructed domain knowledge, including dependency management, build workflows, tool usage, and stage-specific analysis strategies. This knowledge helps the agent adapt its analysis to different projects, but continuously evolving software ecosystems make comprehensive manual maintenance difficult.
A promising direction is to develop self-evolving \textit{SCA-Agent} that learn from previous analyses. By accumulating successful execution paths, failure cases, and dependency evolution patterns, the agent can continuously update its knowledge and optimize planning strategies, reducing manual maintenance and improving adaptability to evolving software ecosystems.

\noindent \textbf{Dataset and Ground-truth Reliability.}
Our dataset is constructed by manually executing the complete software lifecycle of each project and labeling the dependencies observed at each stage. For Runtime, we design execution scripts to exercise major project functionalities and cover dependency loading as comprehensively as possible, while recognizing that observations may still depend on workload coverage. To reduce annotation bias, multiple annotators cross-validate the labels and manually resolve inconsistencies. Dependency resolution may also vary across operating systems, package managers, and execution environments, so we use the same experimental environment for all evaluated methods and consider declared version constraints during component matching to mitigate environment-induced version variations. Although these measures cannot eliminate all ground-truth uncertainty, they improve the consistency and fairness of our evaluation.

\noindent \textbf{Cost--Benefit Trade-off of \textit{SCA-Agent}.}
Compared with traditional SCA tools, \textit{SCA-Agent} introduces additional analysis overhead due to LLM-based reasoning, adaptive Planning, and cross-stage Reasoning. In our experiments, the API cost of \textit{SCA-Agent} is approximately \$0.85 per project, with an average analysis time of about 5 minutes. Most of the overhead comes from project understanding, lifecycle planning, and cross-stage evidence correlation. Although this cost is higher than that of traditional static SCA approaches, \textit{SCA-Agent} is not designed merely to accelerate component scanning, but to improve the accuracy of security assessment through lifecycle context.
Traditional SCA may directly treat vulnerable components observed in intermediate lifecycle stages as risks to the final software. By recovering the actual propagation states of components, \textit{SCA-Agent} improves CVE assessment accuracy by approximately 21.98\%--35\% compared with traditional approaches. For software supply-chain security analysis, more accurate risk assessment can reduce unnecessary investigation and remediation caused by false positives. Therefore, in security-sensitive scenarios, the additional analysis overhead introduced by \textit{SCA-Agent} can be partially offset by the improved quality of security assessment.

\noindent \textbf{Supporting Deeper Supply-chain Security Analysis.}
\textit{SCA-Agent} tracks where a component is introduced, how it propagates across lifecycle stages, and whether it reaches release, deployment, or runtime, allowing vulnerability exposure to be assessed with lifecycle context. The current analysis operates at the component level and does not determine whether vulnerable code is reachable or whether a vulnerability can be triggered during execution. Answering these questions requires additional program-level evidence, such as call relationships, dynamic execution traces, and vulnerability-specific triggering conditions. The provenance, propagation relationships, and runtime observations produced by \textit{SCA-Agent} can be linked with such evidence in future work to extend lifecycle-aware vulnerability assessment toward code reachability and exploitability analysis.

\section{Related Work}
\label{sec:related-work}

Software Composition Analysis (SCA) is a fundamental technique for software supply-chain security, used to identify third-party components, recover dependency relationships, and associate vulnerability information. Early SCA tools mainly relied on project declarations such as manifests and lockfiles for dependency resolution~\cite{eclipse_steady,owasp_dependency_check,ochrona_security}, and their effectiveness has been validated by multiple empirical studies~\cite{yu2024correctness,wu2026more,dann2021identifying,imtiaz2021comparative,zhao2023software}. With the standardization of SBOMs driven by NTIA, SPDX, and CycloneDX~\cite{ntia_sbom_minimum,spdx_specification,owasp_cyclonedx}, tools such as Syft, cdxgen, and Microsoft SBOM Tool further recover components from software packages, file systems, build environments, and container images~\cite{wang2026large,wu2026more,benedetti2025impact}. However, extensive studies have shown that even for the same software object, different tools can produce substantially different component names, versions, and dependency relationships~\cite{yu2024correctness,wang2026large,wu2026more,rabbi2025claim,zhao2025state,garcia2025landscape}, indicating that software composition results are highly dependent on the analyzed representation and evidence source.

To broaden component observability, subsequent studies have explored complementary evidence from different software representations, including incremental SBOM construction~\cite{jia2025sit}, multi-channel analysis for firmware and IoT systems~\cite{safronov2025unibom}, container image analysis~\cite{bufalino2025orca}, and component identification and source matching for stripped or closed-source binaries~\cite{ning2026securing,zhang2025drop,jiang2024binaryai}. Meanwhile, runtime SCA shows that dynamically loaded, reflectively injected, or runtime-resolved components are difficult to capture through static analysis alone. MEM-SBOM and NodeShield complement static evidence through memory forensics and runtime observation, respectively~\cite{alia2026memsbom,cornelissen2025nodeshield}. These studies collectively demonstrate that different software states expose distinct and complementary component evidence, making a single artifact or analysis stage insufficient to characterize software composition.

Recent studies have also begun to compare SCA and SBOM results across lifecycle stages and investigate SBOM consumption consistency, validation, and trustworthy sharing~\cite{rosso2026practical,bufalino2025sbomproof,kishimoto2025dataset,castiglione2026verisbom,sorger2026zksbom}. However, existing approaches mainly acquire evidence independently from specific stages or software objects, without systematically correlating component evidence across its evolution from introduction and build to release, deployment, and runtime. Therefore, recovering complete component lifecycle traces and using them to characterize component origin, propagation, and fate remains a key challenge for lifecycle-aware software composition analysis.
\section{Conclusion}
This paper presents \emph{SCA-Agent}, a lifecycle-aware SCA system that reconstructs component traces across Code, Build, Release, Deploy, and Runtime through adaptive planning and cross-stage reasoning. Evaluation on 105 real-world projects shows that \emph{SCA-Agent} consistently improves component detection and achieves 96.69\% F1 for vulnerability exposure assessment, outperforming the best traditional SCA tool by 18.76 percentage points. These results demonstrate that lifecycle-aware SCA provides a more accurate and traceable foundation for software supply-chain security analysis.

\cleardoublepage

\appendix
\appendix

\section{Open Science}
The prompts and experience files used by \textit{SCA-Agent} are available in our anonymous repository at \url{https://anonymous.4open.science/r/sca_agent-E0AC/}. To support reproducibility, we provide the core prompts, agent configurations, and representative experience files required to understand and reproduce the analysis workflow.

\bibliographystyle{plain}
\bibliography{usenix}

\end{document}